\documentclass{article}
\usepackage[final,nonatbib]{neurips_2024}
\usepackage[utf8]{inputenc} 
\usepackage[T1]{fontenc}    
\usepackage{hyperref}       
\usepackage{url}            
\usepackage{booktabs}       
\usepackage{nicefrac}       
\usepackage{microtype}      
\usepackage{xcolor}         
\usepackage{listings}%
\usepackage{colortbl}
\usepackage{makecell}
\usepackage{subfigure}
\usepackage{graphicx}%
\usepackage{multirow}%
\usepackage{amsmath,amssymb,amsfonts}%
\usepackage{amsthm}%
\usepackage{mathrsfs}%
\usepackage{textcomp}%

\newcommand{\et}{\emph{et al.}\ }

\newcommand{\qfnu}{{\includegraphics[scale=0.035]{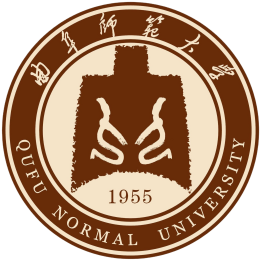}}}
\newcommand{\hit}{{\includegraphics[scale=0.035]{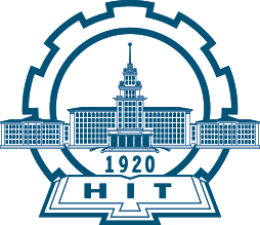}}}

\title{CSWin-UNet: Transformer UNet with Cross-Shaped Windows for Medical Image Segmentation}

\author{
Xiao Liu$^{\qfnu}$, Peng Gao$^{\qfnu}$, Tao Yu$^{\qfnu}$, Fei Wang$^{\hit}$, Ru-Yue Yuan\\
$^{\qfnu}$Qufu Normal University \quad
$^{\hit}$Harbin Institute of Technology Shenzhen
}

\begin{document}

\maketitle

\begin{abstract}
Deep learning, especially convolutional neural networks (CNNs) and Transformer architectures, have become the focus of extensive research in medical image segmentation, achieving impressive results. However, CNNs come with inductive biases that limit their effectiveness in more complex, varied segmentation scenarios. Conversely, while Transformer-based methods excel at capturing global and long-range semantic details, they suffer from high computational demands. In this study, we propose CSWin-UNet, a novel U-shaped segmentation method that incorporates the CSWin self-attention mechanism into the UNet to facilitate horizontal and vertical stripes self-attention. This method significantly enhances both computational efficiency and receptive field interactions. Additionally, our innovative decoder utilizes a content-aware reassembly operator that strategically reassembles features, guided by predicted kernels, for precise image resolution restoration. Our extensive empirical evaluations on diverse datasets, including synapse multi-organ CT, cardiac MRI, and skin lesions, demonstrate that CSWin-UNet maintains low model complexity while delivering high segmentation accuracy. Codes are available at \url{https://github.com/eatbeanss/CSWin-UNet}.
\end{abstract}

\section{Introduction}\label{sec:introduction}

Medical image segmentation is an essential research topic of medical image computing and computer assisted intervention, primarily by processing images to obtain helpful information, such as the shape, size, and structure of diseased organs or tissues, providing more accurate and detailed diagnostic and treatment recommendations\cite{r1,r47}.

Deep learning-based medical image segmentation methods can directly classify the entire image at the pixel level and have been widely applied in multiple medical fields\cite{r2,r3}, including lung computed tomography (CT) image segmentation, brain magnetic resonance image (MRI) segmentation, and cardiac ultrasound image segmentation. These methods not only improve segmentation accuracy but also further advance the field of medical imaging. Convolutional neural network (CNN) is one of the most widely used deep learning technologies in the field of computer vision. The fully convolutional network (FCN)\cite{r4}, an extension of CNN, promotes the development of the field of medical image segmentation. Existing studies have proposed extended convolution and context learning methods to address the limited receptive field of conventional convolutional operations\cite{r40,r41,r42}.
Moreover, UNet\cite{r5}, with its innovative U-shaped encoder-decoder design and skip connections, and merges feature maps from the encoder and decoder to preserve critical spatial details from shallow layers. This architecture has become a staple in image segmentation. Enhanced derivatives of UNet, such as UNet++ \cite{r6}, AttentionUNet\cite{r7}, and ResUNet\cite{r8}, have further refined the segmentation capabilities, offering improved performance across a spectrum of imaging modalities.

Despite the successes of CNN-based methods in medical image segmentation, they are challenged by their limited capacity to capture global and long-range semantic information and inherent inductive biases\cite{r23,r9,r46}. Inspired by the transformative impact of the Transformer architecture in natural language processing (NLP)\cite{r10}, researchers have begun to apply this technology to computer vision tasks, aiming to mitigate some limitations of CNNs\cite{r11,r12,r13}. At the core of the Transformer architecture is the self-attention mechanism, which processes embedded information from all positions within the input sequence in parallel, rather than sequentially. This mechanism allows the Transformer to adeptly manage long-range information dependencies and adapt to varying input sequence lengths. A specific adaptation for image processing, the Vision Transformer\cite{r14}, exemplifies this by segmenting the input image into a series of fixed patches, each converted into a vector that is then processed by a Transformer encoder. Through the encoding stage, self-attention establishes inter-patch relationships, capturing comprehensive contextual information. The resultant encoded features are subsequently utilized in tasks such as object detection and image segmentation, utilizing decoders or classifiers. The introduction of the Vision Transformer has not only infused fresh perspectives into image processing but also achieved results that rival or surpass those of traditional CNNs\cite{r15,r16,r17,r18}. Although the Transformer architecture excels in processing global and long-range semantic information, its computational efficiency is often compromised due to the extensive nature of its self-attention mechanisms. Addressing this inefficiency, the Swin Transformer\cite{r19} innovated with a window self-attention mechanism that limits attention to discrete windows within the image, drastically reducing computational complexity. However, this approach somewhat restricts the interaction among receptive fields. To overcome this, the CSWin Transformer\cite{r20} proposed the cross-shaped window (CSWin) self-attention, which can compute self-attention horizontally and vertically in parallel, achieving better results at a lower computational cost. Additionally, the CSWin Transformer introduced local-enhanced positional encoding (LePE), which imposes positional information on each Transformer block. Unlike previous positional encoding methods\cite{r21,r22}, LePE directly manipulates the results of attention weights rather than being added to the computation of attention. LePE makes the CSWin Transformer more effective for object detection and image segmentation. As the development of Transformers progresses, many studies have combined CNNs with Transformer blocks. TransUNet\cite{r23} and LeViT-UNet\cite{r24} integrated UNet with Transformers and achieved competitive results on abdominal multi-organ and cardiac segmentation datasets. In addition, some researchers have developed segmentation models using pure Transformers. Swin-UNet\cite{r25} employed Swin Transformer blocks to construct encoders and decoders in an UNet-like architecture, showing improved performance compared to TransUNet\cite{r26}. However, this segmentation method based on Swin Transformer still has limitations in receptive field interaction, and the computational cost is also relatively high.

Medical images usually have high resolution and contain many interrelated delicate structures. One of our primary concerns is how to better handle long-range dependencies in medical images with less computational resource consumption. Furthermore, accurate boundary segmentation in medical images is crucial for diagnosis and treatment compared to semantic segmentation. Therefore, another focus of our study is how to retain more detailed information and provide more explicit boundaries during the segmentation process. Inspired by the innovative CSWin Transformer\cite{r20}, we introduce a novel Transformer-based method, dubbed CSWin-UNet, for medical image segmentation. This method is designed to reduce computational costs while simultaneously enhancing segmentation accuracy. Unlike TransUNet\cite{r23}, a CNN-Transformer hybrid architecture, CSWin-UNet, similar to Swin-UNet\cite{r25}, is a pure Transformer-based U-shaped architecture. The critical difference between CSWin-UNet and Swin-UNet is that the former equipped CSWin Transformer blocks in both the encoder and decoder and designed different numbers of blocks according to different scales. Moreover, we introduced the CARAFE (Content-Aware ReAssembly of FEatures) layer\cite{r26} for upsampling in the decoder. Initially, input medical images are transformed into convolutional token embeddings, which are then processed by the encoder to extract contextual features. These features are subsequently upsampled by the CARAFE layer, which enables precise feature reassembly. Additionally, skip connections are employed to fuse high-level semantic information with low-level spatial details continuously. The process culminates in the transformation of feature embeddings into a segmentation mask that matches the original input sizes. Through cross-shaped window self-attention, our method can maintain the efficient feature extraction capability for medical images while reducing computational complexity. Additionally, by combining the classic architecture of UNet, it can effectively integrate features of different scales in the encoder and decoder, thereby improving segmentation accuracy. Finally, the introduction of the CARAFE layer for upsampling can more effectively preserve the edges and detailed features of the segmentation objects. Experimental evaluations of our CSWin-UNet method reveal superior segmentation accuracy and robust generalization capabilities compared to other existing methods. Furthermore, it showcases considerable advantages in reducing computational complexity for medical image segmentation tasks. The key contributions of this study are detailed below:

\begin{itemize}
\item We developed a novel U-shaped encoder-decoder network architecture, CSWin-UNet, utilizing CSWin Transformer blocks specifically tailored for medical image segmentation.
\item The CSWin self-attention mechanism was incorporated to implement horizontal and vertical stripes self-attention learning. This enhancement significantly broadens the focus area for each token, facilitating more comprehensive analysis and contextual integration.
\item In the decoder, the CARAFE layer was employed as an alternative to the traditional transposed convolution or interpolation strategies for upsampling. This choice allows for more accurate pixel-level segmentation masks.
\item Comprehensive experimental results validate that CSWin-UNet is not only lightweight but also demonstrates efficient performance, surpassing existing methods in both computational efficiency and segmentation accuracy.
\end{itemize}

The structure of this paper is organized as follows. Section~\ref{sec:2} reviews recent works and developments in the field of medical image segmentation, setting the context for the innovations introduced in this study. Section~\ref{sec:3} describes in detail the methodology of the newly proposed CSWin-UNet, highlighting the novel aspects of the architecture and its components. Section~\ref{sec:4} presents the experimental results, demonstrating the effectiveness and efficiency of CSWin-UNet compared to existing methods. Section~\ref{sec:5} concludes the paper.

\section{Related works}\label{sec:2}

\subsection{Self-attention mechanisms in image segmentation}

\begin{figure}[t!]
\begin{center}
    \includegraphics[width=0.8\linewidth]{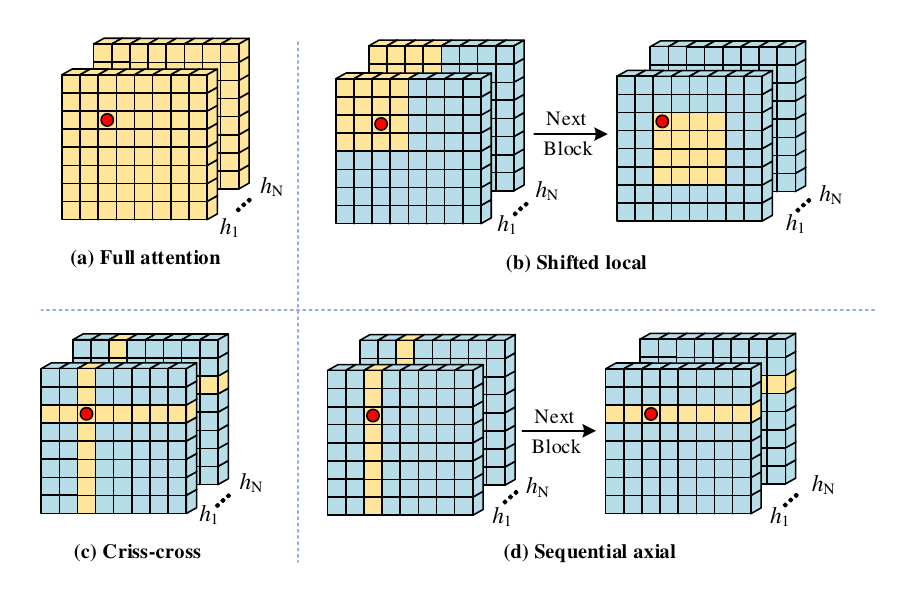}
\end{center}
   \caption{Illustration of different self-attention mechanisms. $h_i$ denotes the $i$-th attention head.}
\label{fig:10}
\end{figure}

The application of self-attention mechanisms in image segmentation has been widely studied. Research in \cite{rd,re,rf} demonstrates that designing different self-attention mechanisms for suitable scenarios can significantly improve segmentation performance. Medical image segmentation tasks often involve subtle but critical structures, and self-attention mechanisms can better capture the relationships between these complex structures, making the design of effective and appropriate self-attention mechanisms particularly important. However, many existing vision Transformers still use global attention mechanisms with high computational complexity, as shown in Fig.\ref{fig:10}(a). To address this issue, Swin Transformer\cite{r19} adopts a shifted version of the local self-attention mechanism, as depicted in Fig.\ref{fig:10}(b), achieving interactions between different windows through a sliding window mechanism. Additionally, axial self-attention\cite{r52} and criss-cross attention\cite{r53} calculate attention within stripes along horizontal and vertical directions, as shown in Fig.\ref{fig:10}(c) and (d), respectively. Nevertheless, axial self-attention is limited by the sequential mechanism and window size, while criss-cross attention performs poorly in specific applications due to overlapping windows. CSWin Transformer\cite{r20} introduces cross-shaped window (CSWin) self-attention, which can compute self-attention for horizontal and vertical stripe regions in parallel. Compared to previous attention mechanisms, this attention mechanism is more general and effective in handling image processing tasks.

\subsection{CNN-based medical image segmentation}

In medical image segmentation, CNNs are predominantly employed, with several vital architectures driving advancements in the field. Among these, FCN\cite{r4} stands out as an end-to-end architecture that classifies pixels directly, converting fully connected layers to convolutional ones to accommodate images of any size. The UNet\cite{r5} model, featuring a symmetric U-shaped encoder-decoder architecture, excels in delivering accurate segmentation of medical images. Building on the foundation laid by FCN and UNet, several enhanced methods have been proposed. For instance, SegNet\cite{r28} incorporates ideas from both FCN and UNet, utilizing max-pooling operators to refine the accuracy of segmentation masks, and has been effectively applied in diverse medical segmentation tasks\cite{r43,r44}. UNet++\cite{r6} extends the original UNet design by integrating densely nested skip connections, which minimize information loss between the encoder and decoder, thereby improving segmentation performance. AttentionUNet\cite{r7} augments the UNet architecture with attention mechanisms to increase both accuracy and robustness. Lastly, nnU-Net\cite{r29} introduces an adaptive approach to network architecture selection, automatically optimizing model configurations based on specific task requirements and dataset characteristics, thus enhancing adaptability across various segmentation challenges. Additionally, MRNet\cite{ra} proposed a multi-rater agreement model to calibrate segmentation results, and Pan \et\cite{rb} designed a hybrid-supervised learning strategy to address the issue of scarce medical image labels.

\subsection{Transformer-based medical image segmentation}

Given the high resolution and complexity of medical images, which encompass a vast number of pixels and intricate local features, traditional CNN-based medical image segmentation methods, while effective at capturing detailed image information, often fall short in accessing global and long-range semantic contexts. In contrast, with their global contextual modeling capabilities, Transformers play a pivotal role in boosting segmentation performance by effectively encoding larger receptive fields and learning relationships between distant pixels. This advantage has spurred researchers to incorporate Transformers into medical image segmentation frameworks. For example, TransUNet\cite{r23} employs a Transformer as the encoder to derive contextual representations from medical images, coupled with a UNet-based decoder for precise pixel-level segmentation. This combination illustrates the enhanced ability of Transformers to capture global contextual information, leading to improved segmentation accuracy. Similarly, TransFuse\cite{r30} integrates CNN and Transformer branches within a single framework, using a specialized module to merge outputs from both pathways to produce final segmentation masks. Further, UNetR\cite{r31} utilizes a Transformer to encode input 3D images, paired with a CNN decoder to complete the segmentation process, while MT-UNet\cite{r32} introduces a hybrid Transformer architecture that learns both intra- and inter-sample relationships. HiFormer\cite{r33} presents another hybrid model, combining two CNNs with a Swin Transformer module and a dual-level fusion module to integrate and transfer multi-scale feature information to the decoder. Among purely Transformer-based methods, Swin-UNet\cite{r25} uses a Swin Transformer\cite{r19} as the encoder to capture global contextual embeddings, which are then progressively upsampled by a UNet decoder, leveraging skip connections to enhance detail preservation. Additionally, DFQ \cite{rc} introduced decoupled feature queries within a Vision Transformer (ViT) framework, enabling segmentation models to adapt more broadly to different tasks.

Inspired by the advancements in multi-head self-attention mechanisms, specifically the CSWin Transformer\cite{r20}, we developed CSWin-UNet, a medical image segmentation method based on CSWin self-attention. This model further conserves computational resources while elevating segmentation accuracy, representing a significant stride forward in the application of Transformers to medical image segmentation.

\section{Methodology}\label{sec:3}

\begin{figure}[t!]
\begin{center}
    \includegraphics[width=0.6\linewidth]{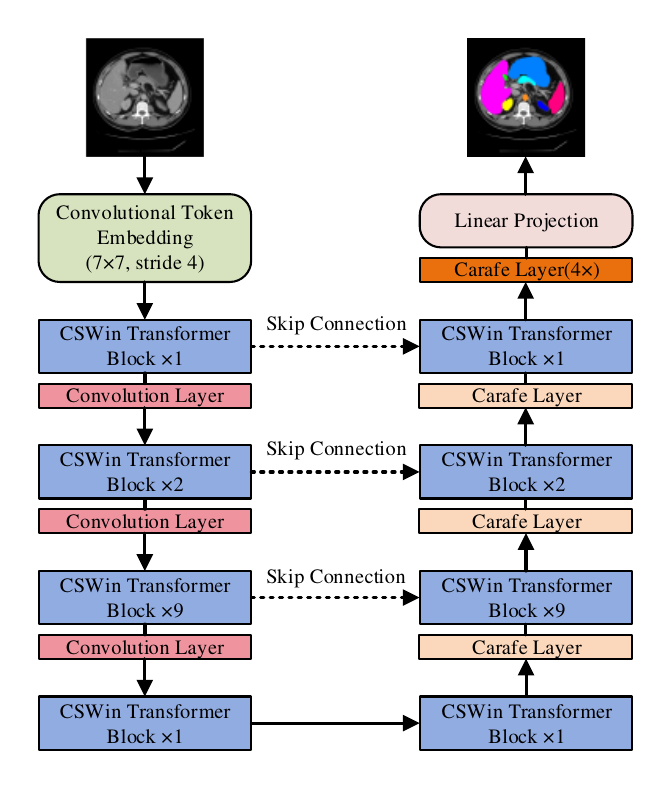}
\end{center}
   \caption{Overview of the proposed CSWin-UNet. The decoder and encoder are symmetrical and each consists of four stages.}
\label{fig:1}
\end{figure}

The overall architecture of CSWin-UNet is illustrated in Fig.\ref{fig:1}, which consists of an encoder, a decoder, and skip connections, with the basic unit being the CSWin Transformer block. For medical images with an input dimension of $H\times W\times 3$, similar to CvT[34], we utilize convolutional token embedding (with a $7\times7$ kernel and a stride of 4) to obtain $H/4\times W/4$ patch tokens with $C$ channels. Both the encoder and decoder are consisted of four stages. Like UNet[5], skip connections are employed to merge features at each stage of the encoder and decoder to retain contextual information better. In the encoder, convolutional layers (with a $3\times3$ kernel and a stride of 2) are used for downsampling, reducing the resolution to half of its input size while doubling the channel count. Upsampling in the decoder is performed through the CARAFE Layer, increasing the resolution to twice its input size while halving the channel count. Finally, a $4\times$ CARAFE upsampling operation is performed to restore the resolution to the input resolution $H\times W$, and a linear layer is used to convert the feature map into a segmentation mask.

\subsection{CSWin Transformer block}

\begin{figure}[t!]
\begin{center}
    \includegraphics[width=0.2\linewidth]{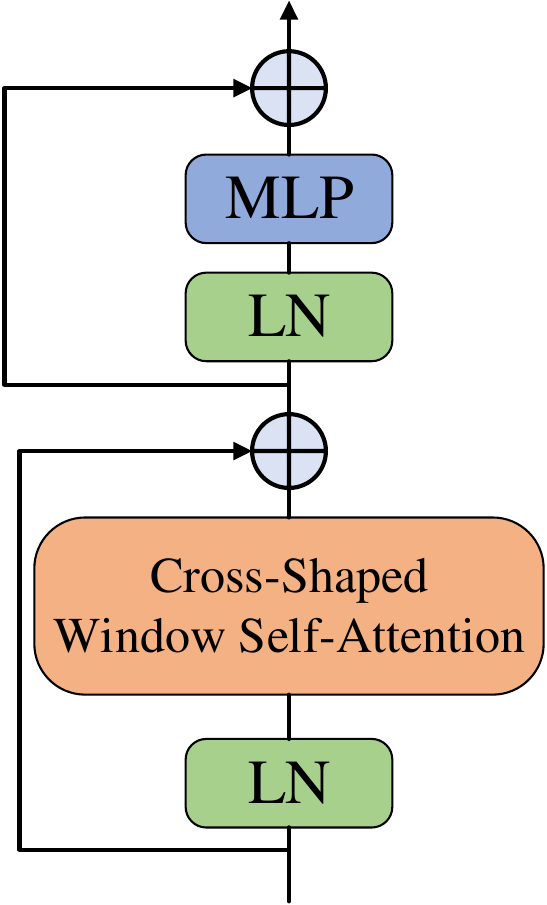}
\end{center}
   \caption{Pipeline of the CSWin Transformer Block.}
\label{fig:2}
\end{figure}

With its self-attention mechanism, the traditional Transformer architecture excels in establishing global semantic dependencies by processing all pixel positions, which, however, leads to high computational costs in high-resolution medical imaging. The Swin Transformer\cite{r19} mitigates these costs with a shifted window attention mechanism that divides the image into distinct, non-overlapping windows, allowing for localized self-attention. This adaptation helps manage the high resolution of images while controlling computational complexity. Yet, the effectiveness of this approach depends on the window size; smaller windows might miss some global information, while larger ones could unnecessarily raise computational demands and storage. Contrasting with the shifted window attention mechanism, CSWin self-attention organizes attention into horizontal and vertical stripes, enhancing parallel computation capabilities. This structure not only conserves computational resources but also broadens the interaction within receptive fields. As depicted in Fig.\ref{fig:2}, the CSWin Transformer block, built on this innovative self-attention design, includes a CSWin self-attention module, a LayerNorm (LN) layer, a multi-layer perceptron (MLP), and skip connections. This configuration optimally balances local and global information processing, significantly improving efficiency and effectiveness in complex medical image segmentation tasks.

\begin{figure}[t!]
\begin{center}
    \includegraphics[width=0.6\linewidth]{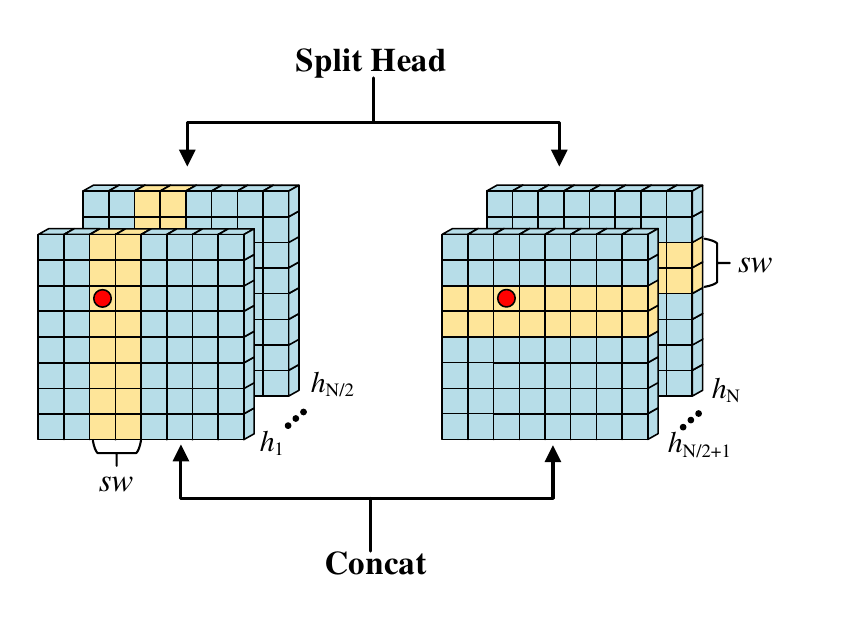}
\end{center}
   \caption{Illustration of the CSWin self-attention mechanism. First, split the multiple heads $\{h_1,h_2,\ldots,h_N\}$ into two groups $\{h_1,h_2,\ldots,h_{N/2}\}$ and $\{h_{N/2+1},h_{N/2+2},\ldots,h_N\}$, performing self-attention in parallel on the horizontal and vertical stripes, respectively, and concatenate the outputs. Next, the width of the stripe $sw$ can be adjusted to achieve optimal performance. Generally, choose a smaller $sw$ for higher resolutions and a larger $sw$ for lower resolutions.}
\label{fig:3}
\end{figure}

In the multi-head self-attention mechanism, the input feature $X \in \mathbb{R}^{H \times W \times C}$ undergoes an initial transformation where it is linearly mapped across $N$ heads, with $N$ typically chosen as an even number. Distinct from conventional self-attention and the shifted window-based multi-head self-attention, the CSWin self-attention uniquely facilitates local self-attention learning within divided horizontal or vertical stripes, as depicted in Fig.\ref{fig:3}. This configuration allows each head to compute self-attention within its designated stripe either horizontally or vertically. These operations are performed in parallel, effectively broadening the scope of the attention computation area while simultaneously reducing the overall computational complexity.

In the horizontal stripes self-attention configuration of the CSWin Transformer, the input feature $X$ is systematically divided into $M$ non-overlapping horizontal stripes, represented as $[X_h^1, X_h^2,\ldots, X_h^M]$, where each stripe has a width $sw$, and $M$ is determined by the ratio $H/sw$. The parameter $sw$ is adjustable and critical for balancing computational complexity with the model's learning capability. Specifically, a larger $sw$ enhances the model's ability to explore long-range pixel correlations within each stripe, potentially capturing more extensive contextual information. Consider the computation within a specific head, denoted as the $n$-th head. In this scenario, the dimensions of the queries (Q), keys (K), and values (V) are each $d_n=C/N$, where $C$ is the number of channels, and $N$ is the total number of heads. The self-attention output $Y_n^i$ for the $n$-th head within the $i$-th horizontal stripe is calculated as follows:
\begin{equation}\label{eq:1}
  \begin{aligned}
    Q_n^i & =W^Q_nX_h^i \\
    K_n^i & =W^K_nX_h^i \\
    V_n^i & =W^V_nX_h^i \\
    Y_n^i &=\text{Softmax}\left(\frac{Q_n^i(K_n^i)^T}{\sqrt{d_n}}\right)V_n^i
  \end{aligned}
\end{equation}
where $X_h^i\in\mathbb{R}^{sw\times W\times C}$ is the feature map of the $i$-th horizontal stripe; $W^Q_n\in\mathbb{R}^{C\times d_n}$, $W^K_n\in\mathbb{R}^{C\times d_n}$, and $W^V_n\in\mathbb{R}^{C\times d_n}$ represent the weight matrix of Q, K and V of the $n$-th head. This operation is performed separately and in parallel for each stripe to allow self-attention within that specific horizontal stripe. The self-attention from $M$ horizontal stripes are concatenated to construct the horizontal self-attention $\text{H-Attention}_n$ for the $n$-th head:
\begin{equation}\label{eq:2}
  \begin{aligned}
    &X=[X_h^1,X_h^2,\ldots,X_h^M]\\
    &\text{H-Attention}_n(X)=[Y_n^1,Y_n^2,\ldots,Y_n^M]
  \end{aligned}
\end{equation}

Similar to horizontal stripes self-attention, the input feature $X\in\mathbb{R}^{H\times W\times C}$ is evenly divided into $S$ non-overlapping vertical stripes $[X_v^1,X_v^2,\ldots,X_v^S]$ for vertical self-attention, where the stripe width is also $sw$, and $S=W/sw$. Considering the $n$-th attention head as an example, where the dimensions of Q, K, and V are $d_n=C/N$. The self-attention output $Y_n^i$ for the $n$-th head within the $i$-th vertical stripe can be computed as follows:
\begin{equation}\label{eq:3}
  \begin{aligned}
    Q_n^i & =W^Q_nX_v^i \\
    K_n^i & =W^K_nX_v^i \\
    V_n^i & =W^V_nX_v^i \\
    Y_n^i &=\text{Softmax}\left(\frac{Q_n^i(K_n^i)^T}{\sqrt{d_n}}\right)V_n^i
  \end{aligned}
\end{equation}
where $X_v^i\in\mathbb{R}^{H\times sw\times C}$ is the feature map of the $i$-th vertical stripe. The self-attention from $S$ vertical stripes are concatenated to construct the vertical self-attention $\text{V-Attention}_n$ for the $n$-th head:
\begin{equation}\label{eq:4}
  \begin{aligned}
    &X=[X_v^1,X_v^2,\ldots,X_v^S]\\
    &\text{V-Attention}_n(X)=[Y_n^1,Y_n^2,\ldots,Y_n^S]
  \end{aligned}
\end{equation}

We split the $N$ heads into two groups, each containing $N/2$ heads. Each head within these groups generates its self-attention output. The first group is tasked with learning horizontal stripes self-attention, while the second group learns vertical stripes self-attention. After computing the self-attention separately, the outputs from these two groups are concatenated. This concatenation is performed along the channel dimension as:
\begin{equation}\label{eq:5}
  \begin{aligned}
    &\text{h}_n=\left\{
        \begin{aligned}
            \text{H-Attention}_n(X)&,\quad n=1,2,\ldots,N/2 \\
            \text{V-Attention}_n(X)&,\quad n=N/2+1,N/2+2,\ldots,N
        \end{aligned}
    \right. \\
    &\text{CSWin-Attention}(X)=\mathrm{concat}(\text{h}_1,\text{h}_2,\ldots,\text{h}_N)W^o
  \end{aligned}
\end{equation}
where $\text{h}_n$ denotes the $n$-th attention head; $W^o\in\mathbb{R}^{C\times C}$ is a weight matrix used to linearly transform the concatenated output of the multi-head self-attention mechanism to produce the final attention output, this linear transformation can help to learn the relationship between different heads and fuse the attention information. The concatenated output effectively incorporates horizontal and vertical contextual information, comprehensively learning the spatial relationships within the input image.

Based on the above self-attention mechanism, the CSWin Transformer block can be defined as:
\begin{equation}\label{eq:6}
  \begin{aligned}
    \hat{X}^l&=\text{CSWin-Attention}\left(\text{LN}(X^{l-1})\right)+X^{l-1} \\
    X^l&=\text{MLP}\left(\text{LN}(\hat{X}^l)\right)+\hat{X}^l
  \end{aligned}
\end{equation}
where $X^l$ represents the output of the $l$-th CSWin Transformer block or the precedent convolutional layer of each stage.

\subsection{Encoder}

In the encoder, the input image dimensions are $H/4\times W/4\times C$, which then enter four stages for feature extraction. Downsampling operations accompany the first three stages. The number of CSWin Transformer blocks varies across the four stages, and details on block count settings will be discussed later. The downsampling layer is implemented by a convolutional layer with a kernel size of $3\times3$ and a stride of 2, which reduces the resolution to half of its input size while doubling the number of channels. The stripe width $sw$ changes according in different stages. As the resolution continuously decreases and the channel number increases, a smaller $sw$ is chosen in stages with larger resolutions and a larger $sw$ in stages with smaller resolutions, effectively enlarging the attention region of each token in stages with smaller resolutions. Additionally, the input image resolution is $224\times 224$. To ensure the mediate feature map size of the input image can be divided by $sw$, we set $sw$ for the four stages are 1, 2, 7, and 7.

\subsection{Decoder}

Corresponding to the encoder, the decoder also has four stages. Image resolution and channel number increases are achieved through the CARAFE layer in the last three stages. The number of CSWin Transformer blocks and the stripe width $sw$ for attention learning in the four stages are consistent with those set in the encoder. Commonly used upsampling methods include linear interpolation and transposed convolution. Bilinear interpolation only considers adjacent pixels and may blur the edges of the image, leading to unclear boundaries in the segmentation results, whereas the receptive field of transposed convolution is usually constrained by the kernel size and stride, not only limiting its ability to represent local variations but also requiring the learning of the weights and biases of the transposed convolution kernels. Unlike these methods, we use the CARAFE\cite{r26} to achieve upsampling.

The CARAFE layer is an advanced upsampling mechanism comprising two principal components: a kernel prediction module and a content-aware reassembly module. The kernel prediction module initiates with a convolution layer tasked with predicting the reassembly kernels from the encoded features. It includes three sub-modules: channel compressor, context encoder, and kernel normalizer. The channel compressor reduces the dimensionality of the channel space in the input feature map $X\in\mathbb{R}^{H\times W\times C}$, thereby lowering computational complexity and focusing on essential feature information. Following channel compression, the context encoder processes the reduced feature map to encode contextual information, which is instrumental in generating the reassembly kernels. Each predicted reassembly kernel undergoes normalization through a Softmax function in the kernel normalizer to ensure that the output distribution of the weights is probabilistic and sums to one, enhancing the stability and performance of the upsampling process. With an upsampling ratio $\sigma$ (where $\sigma$ is an integer), CARAFE aims to generate an expanded feature map $X'\in\mathbb{R}^{\sigma H\times \sigma W\times C}$. For each pixel $l'=(i', j')$ in $X'$, it corresponds to a specific pixel $l=(i,j)$ in $X$, determined by $i=\lfloor i'/d\rfloor$ and $j=\lfloor j'/d\rfloor$. The kernel prediction module$\psi$ predicts a unique reassembly kernel $W_{l'}$ for each pixel $l'$ based on the neighborhood $\mathcal{N}(X_l,k)$, which is a $k\times k$ region centered around pixel $l$ on $X$. This neighborhood extracts localized features, which the predicted kernel uses to effectively reassemble and upsample the feature map.
\begin{equation}\label{eq:7}
  W_{l'}=\psi\left(\mathcal{N}(X_l,k_{\text{encoder}})\right)
\end{equation}
where $k_{\text{encoder}}$ denotes the receptive filed of the content encoder.

The second step is content-aware reassembly, where the input features are reassembled using a convolutional layer, and the content-aware reassembly module $\phi$ reassembles $\mathcal{N}(X_l,k)$ with the reassembly kernel $W_{l'}$:
\begin{equation}\label{eq:8}
  X'_{l'}=\phi\left(\mathcal{N}(X_l,k_{\text{up}}),W_{l'}\right)
\end{equation}
where $k_{\text{up}}$ is the size of reassembly kernel. For each reassembly kernel $W_{l'}$, the content-aware reassembly module reassembles the features within the local square region. The module $\phi$ performs a weighted summation. For a pixel position $l$ and its centered neighborhood $\mathcal{N}(X_l,k_{\text{up}})$, the reassembly process is as follows:
\begin{equation}\label{eq:9}
  X'_{l'}=\sum_{n=-r}^{r}\sum_{m=-r}^{r}W_{l'(n,m)}\cdot X_{(i+n,j+m)}
\end{equation}
where $r=\lfloor k_{\text{up}}/2 \rfloor$.

Each pixel within $\mathcal{N}(X_l,k_{\text{up}})$ contributes differently to the upsampled pixel $l'$. The reassembled feature map can enhance the focus on relevant information within the local area, providing more robust semantic information than the original feature map. Additionally, similar to UNet\cite{r5}, we use skip connections to merge the feature maps outputted from the encoder and decoder, thereby providing more prosperous and more accurate spatial information that helps recover image detail. Subsequently, a $1\times 1$ convolution kernel is used to reduce the number of channels after concatenation, ensuring consistency with the number of feature channels in the upsampling process.

\section{Experiments}\label{sec:4}

\subsection{Implementation details}

CSWin-UNet is implemented using Python and the PyTorch framework. The model training and evaluation are conducted on an NVIDIA$^\circledR$ GeForce RTX$^{\mathrm{TM}}$ 3090 GPU with 24GB VRAM. We initialize the CSWin Transformer blocks with pretrained weights from ImageNet\cite{r38} to leverage prior knowledge and accelerate the convergence process. For data augmentation, schemes such as flipping and rotation are employed to enhance the diversity of the training dataset, thereby helping the model generalize better to unseen data. During the training phase, the batch size is set to 24, and we use a learning rate of 0.05. Optimization is performed using the stochastic gradient descent (SGD) method with a momentum of 0.9 and a weight decay factor of $10^{-4}$. This setup is chosen to optimize the balance between rapid learning and convergence stability. Furthermore, to effectively train CSWin-UNet, we employ a combined loss function that integrates Dice and cross-entropy losses, defined as follows:
\begin{equation}\label{eq:10}
  Loss=\alpha Loss_{Dice}+\beta Loss_{cross}
\end{equation}
where $\alpha$ and $\beta$ are two hyperparameters used to balance the impact of $Loss_{Dice}$ and $Loss_{cross}$ on the final loss, respectively. This combined loss targets both pixel-level accuracy and holistic segmentation quality, ensuring robust learning and improved performance across varied medical image segmentation tasks.

\subsection{Datasets and metrics}

\subsubsection{Synapse dataset}
The synapse multi-organ segmentation dataset includes 30 CT scans from the MICCAI 2015 Multi-Atlas Abdominal Organ Segmentation Challenging, encompassing a total of 3779 abdominal CT images. Each CT scan consists of 85 to 198 slices with $512\times512$ pixels each, and each voxel measures $([0.54, 0.54]\times[0.98, 0.98]\times[2.5, 5.0]) \text{mm}^3$. Following the settings in literature\cite{r37,r23}, 18 sets are selected for training and 12 for evaluation. The segmentation performance on eight types of abdominal organs (aorta, gallbladder, left kidney, right kidney, liver, pancreas, spleen, stomach) is evaluated using the mean Dice-similarity coefficient (DSC) and the mean Hausdorff Distance (HD) as metrics.

\subsubsection{ACDC dataset}

The Automated Cardiac Diagnosis Challenge (ACDC) dataset was released during the 2017 ACDC challenge and raised a multi-category cardiac 3D MRI dataset that includes 100 sets of short-axis MR cardiac images obtained through cine MR 1.5T and 3T scanners. Medical experts provided annotations for three cardiac structures: the right ventricle (RV), myocardium (MYO), and left ventricle (LV) \cite{r45}. We randomly selected 70 sets of MR images for training, 10 for validation, and 20 for evaluation. The ACDC dataset uses the mean DSC as an evaluation metric to assess the segmentation results of the three cardiac structures.

\subsubsection{Skin lesion segmentation datasets}

We conducted experiments on the ISIC2017\cite{r48}, ISIC2018\cite{r49}, and PH$^2$\cite{r50} datasets. The ISIC datasets include a large number of dermoscopic images, covering various skin lesions. Following the settings in HiFormer\cite{r33}, we used 1400 images for training, 200 images for validation, and 400 images for testing in the ISIC2017 dataset; 1815 images for training, 259 images for validation, and 520 images for testing in the ISIC2018 dataset; and 80 images for training, 20 images for validation, and 100 images for testing in the PH$^2$ dataset. We evaluated the skin lesion segmentation task using mean DSC, sensitivity (SE), specificity (SP), and accuracy (ACC) as metrics.

\subsection{Results on Synapse dataset}

\begin{table}[t!]
\centering
\caption{Detailed comparisons with recent medical image segmentation methods in terms of DSC and HD on the Synapse dataset. The first and second best values are highlighted in \textbf{\textcolor[rgb]{1.00,0.00,0.00}{red}} and \textbf{\textcolor[rgb]{0.00,0.00,1.00}{blue}} fonts, respectively.}
\label{tab:1}
\resizebox{\textwidth}{!}{%
\begin{tabular}{l|c|cc|cccccccc}
\toprule
Methods   & Backbone & DSC $\uparrow$  & HD $\downarrow$   & Aorta & Gallbladder & Kidney (L) & Kidney (R) & Liver & Pancreas & Spleen & Stomach \\
\midrule
V-Net\cite{r36} & CNN     & 68.81 & -     & 75.34 & 51.87       & 77.10     & \textbf{\textcolor[rgb]{1.00,0.00,0.00}{80.75}}     & 87.84 & 40.05    & 80.56  & 56.98   \\
DARR\cite{r37} & CNN      & 69.77 & -     & 74.74 & 53.77       & 72.31     & 73.24     & 94.08 & 54.18    & 89.90  & 45.96   \\
UNet\cite{r5} & CNN      & 76.85 & 39.70 & \textbf{\textcolor[rgb]{0.00,0.00,1.00}{89.07}} & \textbf{\textcolor[rgb]{1.00,0.00,0.00}{69.72}}       & 77.77     & 68.60     & 93.43 & 53.98    & 86.67  & 75.58   \\
Att-UNet\cite{r7} & CNN  & 77.77 & 36.02 & \textbf{\textcolor[rgb]{1.00,0.00,0.00}{89.55}} & \textbf{\textcolor[rgb]{0.00,0.00,1.00}{68.88}}       & 77.98     & 71.11     & 93.57 & 58.04    & 87.30  & 75.75   \\
HiFormer-B\cite{r33} & CNN+Transformer & \textbf{\textcolor[rgb]{0.00,0.00,1.00}{80.39}} & \textbf{\textcolor[rgb]{1.00,0.00,0.00}{14.70}} & 86.21 & 65.69       & \textbf{\textcolor[rgb]{1.00,0.00,0.00}{85.23}}     & 79.77     & \textbf{\textcolor[rgb]{0.00,0.00,1.00}{94.61}} & 59.52    & \textbf{\textcolor[rgb]{1.00,0.00,0.00}{90.99}}  & \textbf{\textcolor[rgb]{0.00,0.00,1.00}{81.08}}   \\
MT-UNet\cite{r32} & CNN+Transformer   & 78.59 & 26.59 & 87.92 & 64.99       & 81.47     & 77.29     & 93.06 & \textbf{\textcolor[rgb]{0.00,0.00,1.00}{59.46}}    & 87.75  & 76.81   \\
TransUNet\cite{r23} & CNN+Transformer & 77.48 & 31.69 & 87.23 & 63.13       & 81.87     & 77.02     & 94.08 & 55.86    & 85.08  & 75.62   \\
Swin-UNet\cite{r25} & Transformer & 79.13 & 21.55 & 85.47 & 66.53       & 83.28     & \textbf{\textcolor[rgb]{0.00,0.00,1.00}{79.61}}     & 94.29 & 56.58    & \textbf{\textcolor[rgb]{0.00,0.00,1.00}{90.66}}  & 76.60   \\
\hline
CSWin-UNet & Transformer & \textbf{\textcolor[rgb]{1.00,0.00,0.00}{81.12}} & \textbf{\textcolor[rgb]{0.00,0.00,1.00}{18.86}} & 87.13 & 67.85       & \textbf{\textcolor[rgb]{0.00,0.00,1.00}{83.51}}     & 78.53     & \textbf{\textcolor[rgb]{1.00,0.00,0.00}{95.23}} & \textbf{\textcolor[rgb]{1.00,0.00,0.00}{65.94}}    & 89.05  & \textbf{\textcolor[rgb]{1.00,0.00,0.00}{81.74}}    \\
\bottomrule
\end{tabular}%
}
\end{table}

\begin{figure}[t]
\begin{center}
    \includegraphics[width=0.8\linewidth]{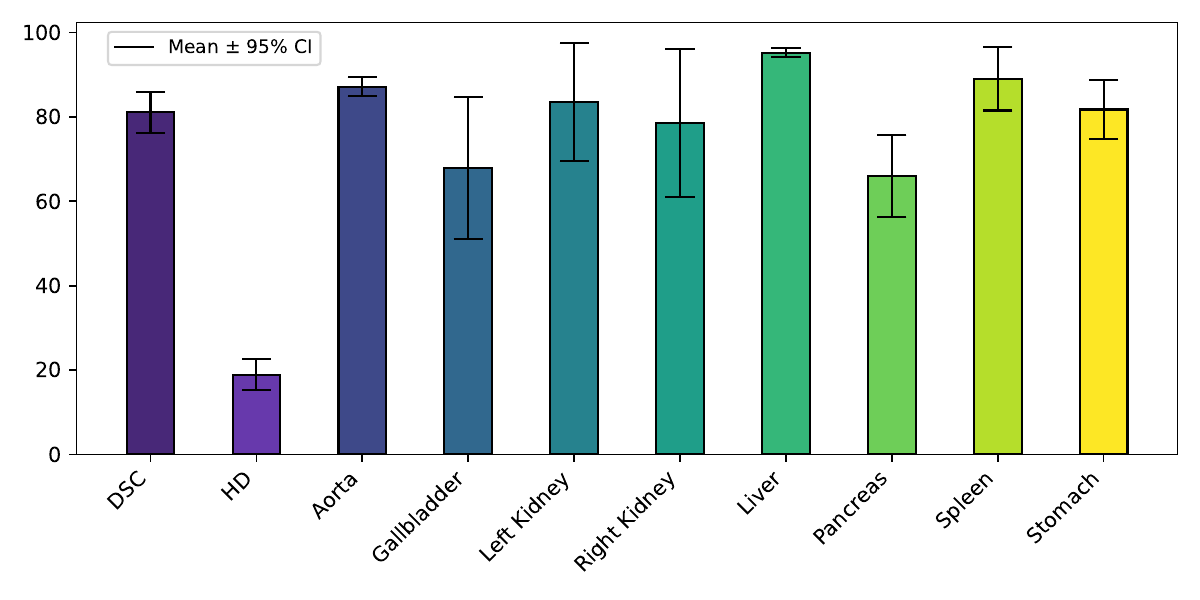}
\end{center}
   \caption{Error bars (95\% confidence interval) of mean DSC, mean HD, and DSC for each organ on the Synapse dataset.}
\label{fig:5}
\end{figure}

As shown in Table \ref{tab:1}, our proposed method on the Synapse dataset has improved the mean DSC and HD. Meanwhile, we represent error bars (95\% confidence interval) of mean DSC, mean HD, and DSC for each organ in Fig.\ref{fig:5}. Compared to TransUNet\cite{r23} and Swin-UNet\cite{r25}, our mean DSC has increased by 3.64\% and 1.99\%, respectively, and the mean HD has improved by 12.83\% and 2.69\%. Notably, in the segmentation of the pancreas, the DSC of CSWin-UNet is significantly higher than that of other segmentation methods. Unlike other organs, the pancreas has blurry boundaries and variability, and our method achieved more precise segmentation results for the pancreas, indicating that our CSWin-UNet provides higher segmentation accuracy in complex segmentation environments.

\begin{figure}[t!]
\begin{center}
    \includegraphics[width=\linewidth]{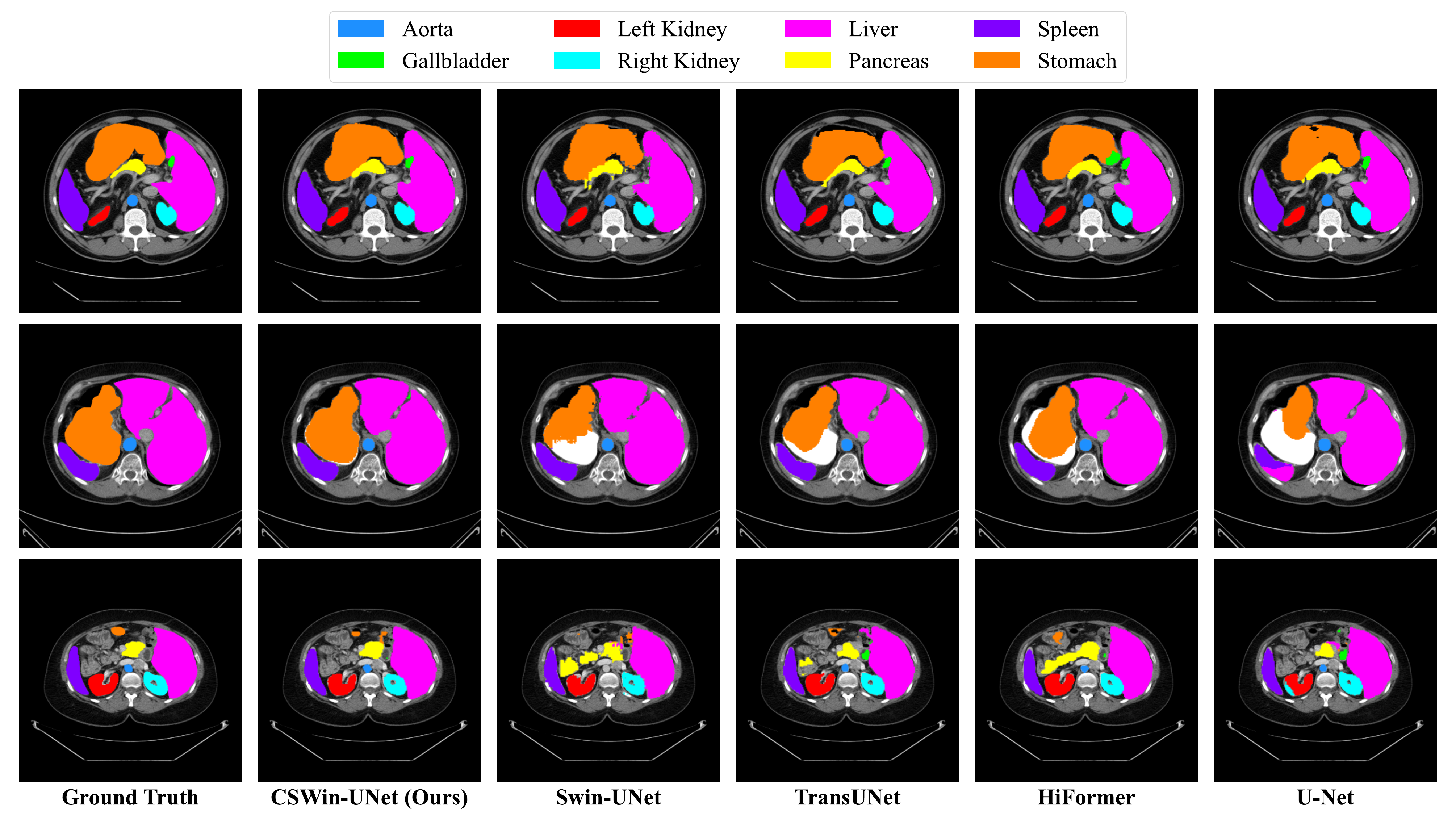}
\end{center}
   \caption{Segmentation results of eight abdominal organs (aorta, gallbladder, left kidney, right kidney, liver, pancreas, spleen, stomach) in the Synapse dataset using different methods.}
\label{fig:4}
\end{figure}

To more intuitively evaluate the proposed method, we conducted a visual analysis of the segmentation results. Fig.\ref{fig:4} displays the comparative results on the Synapse dataset. The first row shows that Swin-UNet and HiFormer-B exhibit significant errors in segmenting small organs like the gallbladder (green label), with Swin-UNet\cite{r25} failing to delineate boundaries accurately and HiFormer-B\cite{r33} mistakenly identifying other areas as the gallbladder. The second row indicates that Swin-UNet, TransUNet\cite{r23}, HiFormer-B, and UNet\cite{r5} all fail to segment the stomach (orange label) completely. The third row reveals that Swin-UNet and HiFormer-B incorrectly label large areas of other organs as the pancreas (yellow label). Considering the quantitative metrics and visual results, our proposed CSWin-UNet achieves accurate segmentation of refined and complex organs, produces more accurate segmentation results, demonstrates greater robustness against complex backgrounds, and performs superior edge structure handling.

\subsection{Results on ACDC dataset}

\begin{table}[t!]
\centering
\caption{Comparison results of top-tier medical image segmentation methods on the ACDC dataset. The first and second best values are highlighted in \textbf{\textcolor[rgb]{1.00,0.00,0.00}{red}} and \textbf{\textcolor[rgb]{0.00,0.00,1.00}{blue}} fonts, respectively.}
\label{tab:2}
\begin{tabular}{l|c|c|ccc}
\toprule
Methods    & Backbone & DSC $\uparrow$  & RV    & MYO   & LV    \\
\midrule
UNet\cite{r5} & CNN     & 87.55 & 87.10 & 80.63 & 94.92 \\
Att-UNet\cite{r7} & CNN  & 86.75 & 87.58 & 79.20 & 93.47 \\
nnUNet\cite{r29} & CNN    & \textbf{\textcolor[rgb]{0.00,0.00,1.00}{90.91}} & \textbf{\textcolor[rgb]{0.00,0.00,1.00}{89.21}} & \textbf{\textcolor[rgb]{1.00,0.00,0.00}{90.20}} & 93.35 \\
UNetR\cite{r31} & CNN+Transformer     & 88.61 & 85.29 & 86.52 & 94.02 \\
TransUNet\cite{r23} & CNN+Transformer & 89.71 & 88.86 & 84.53 & 95.73 \\
Swin-UNet\cite{r25} & Transformer & 90.00 & 88.55 & 85.62 & \textbf{\textcolor[rgb]{1.00,0.00,0.00}{95.83}} \\
\hline
CSWin-UNet & Transformer & \textbf{\textcolor[rgb]{1.00,0.00,0.00}{91.46}} & \textbf{\textcolor[rgb]{1.00,0.00,0.00}{89.68}} & \textbf{\textcolor[rgb]{0.00,0.00,1.00}{88.94}} & \textbf{\textcolor[rgb]{0.00,0.00,1.00}{95.76}} \\
\bottomrule
\end{tabular}%
\end{table}

\begin{figure}[t!]
\begin{center}
    \includegraphics[width=0.4\linewidth]{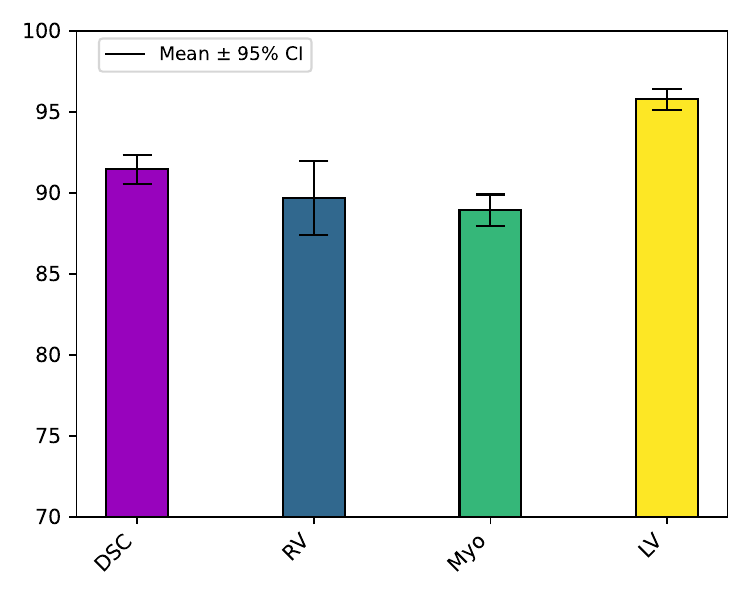}
\end{center}
   \caption{Error bars (95\% confidence interval) of mean DSC and DSC for each cardiac structure on the ACDC dataset.}
\label{fig:6}
\end{figure}

Table \ref{tab:2} presents the experimental results of our proposed CSWin-UNet on the ACDC dataset and compares them with other advanced methods. Fig.\ref{fig:6} represents the error bars (95\% confidence interval) of mean DSC and DSC for each cardiac structure. In the table, RV represents the right ventricle, MYO represents the myocardium, and LV represents the left ventricle. The results indicate that the proposed CSWin-UNet achieves better identification and segmentation of these organs, with an accuracy rate of 91.40\%, demonstrating good generalization capabilities and robustness.

\subsection{Results on skin lesion segmentation datasets}

\begin{table}[t!]
\centering
\caption{Comparative results of the proposed CSWin-UNet with other state-of-the-art methods on three skin lesion segmentation datasets. The first and second best values are highlighted in \textbf{\textcolor[rgb]{1.00,0.00,0.00}{red}} and \textbf{\textcolor[rgb]{0.00,0.00,1.00}{blue}} fonts, respectively.}
\label{tab:7}
\resizebox{\textwidth}{!}{%
\begin{tabular}{l|c|cccc|cccc|cccc}
\toprule
\multirow{2}{*}{Method} & \multirow{2}{*}{Backbone} & \multicolumn{4}{c|}{ISIC2017\cite{r48}}  & \multicolumn{4}{c|}{ISIC2018\cite{r49}}  & \multicolumn{4}{c}{PH$^2$\cite{r50}}       \\ \cline{3-14}
                        & & DSC $\uparrow$  & SE $\uparrow$   & SP $\uparrow$   & ACC $\uparrow$  & DSC $\uparrow$  & SE $\uparrow$   & SP $\uparrow$   & ACC $\uparrow$  & DSC $\uparrow$  & SE $\uparrow$   & SP $\uparrow$   & ACC $\uparrow$  \\
\hline
UNet\cite{r5}          & CNN & 81.59 & 81.72 & 96.80 & 91.64 & 85.45 & 88.00 & 96.97 & 94.04 & 89.36 & 91.25 & 95.88 & 92.33 \\
Att-UNet\cite{r7}       & CNN & 80.82 & 79.98 & 97.76 & 91.45 & 85.66 & 86.74 & \textbf{\textcolor[rgb]{1.00,0.00,0.00}{98.63}} & 93.76 & 90.03 & 92.05 & 96.40 & 92.76 \\
TransUNet\cite{r23}     & CNN+Transformer & 81.23 & 82.63 & 95.77 & 92.07 & 84.99 & 85.78 & 96.53 & 94.52 & 88.40 & 90.63 & 94.27 & 92.00 \\
HiFormer-B\cite{r33}    & CNN+Transformer & \textbf{\textcolor[rgb]{1.00,0.00,0.00}{92.53}} & \textbf{\textcolor[rgb]{0.00,0.00,1.00}{91.55}} & \textbf{\textcolor[rgb]{0.00,0.00,1.00}{98.40}} & \textbf{\textcolor[rgb]{0.00,0.00,1.00}{97.02}} & \textbf{\textcolor[rgb]{0.00,0.00,1.00}{91.02}} & \textbf{\textcolor[rgb]{0.00,0.00,1.00}{91.19}} & 97.55 & \textbf{\textcolor[rgb]{0.00,0.00,1.00}{96.21}} & \textbf{\textcolor[rgb]{1.00,0.00,0.00}{94.60}} & \textbf{\textcolor[rgb]{0.00,0.00,1.00}{94.20}} & \textbf{\textcolor[rgb]{0.00,0.00,1.00}{97.72}} & 96.61 \\
Swin-UNet\cite{r25}     & Transformer & \textbf{\textcolor[rgb]{0.00,0.00,1.00}{91.83}} & 91.42 & 97.98 & 97.01 & 89.46 & 90.56 & \textbf{\textcolor[rgb]{0.00,0.00,1.00}{97.98}} & \textbf{\textcolor[rgb]{1.00,0.00,0.00}{96.45}} & \textbf{\textcolor[rgb]{0.00,0.00,1.00}{94.49}} & 94.10 & 95.64 & \textbf{\textcolor[rgb]{0.00,0.00,1.00}{96.78}} \\
\hline
CSWin-UNet              & Transformer & 91.47 & \textbf{\textcolor[rgb]{1.00,0.00,0.00}{93.79}} & \textbf{\textcolor[rgb]{1.00,0.00,0.00}{98.56}} & \textbf{\textcolor[rgb]{1.00,0.00,0.00}{97.26}} & \textbf{\textcolor[rgb]{1.00,0.00,0.00}{91.11}} & \textbf{\textcolor[rgb]{1.00,0.00,0.00}{92.31}} & 97.88 & 95.25 & 94.29 & \textbf{\textcolor[rgb]{1.00,0.00,0.00}{95.63}} & \textbf{\textcolor[rgb]{1.00,0.00,0.00}{97.82}} & \textbf{\textcolor[rgb]{1.00,0.00,0.00}{96.82}} \\
\bottomrule
\end{tabular}%
}
\end{table}

\begin{figure}[t!]
\begin{center}
    \includegraphics[width=0.8\linewidth]{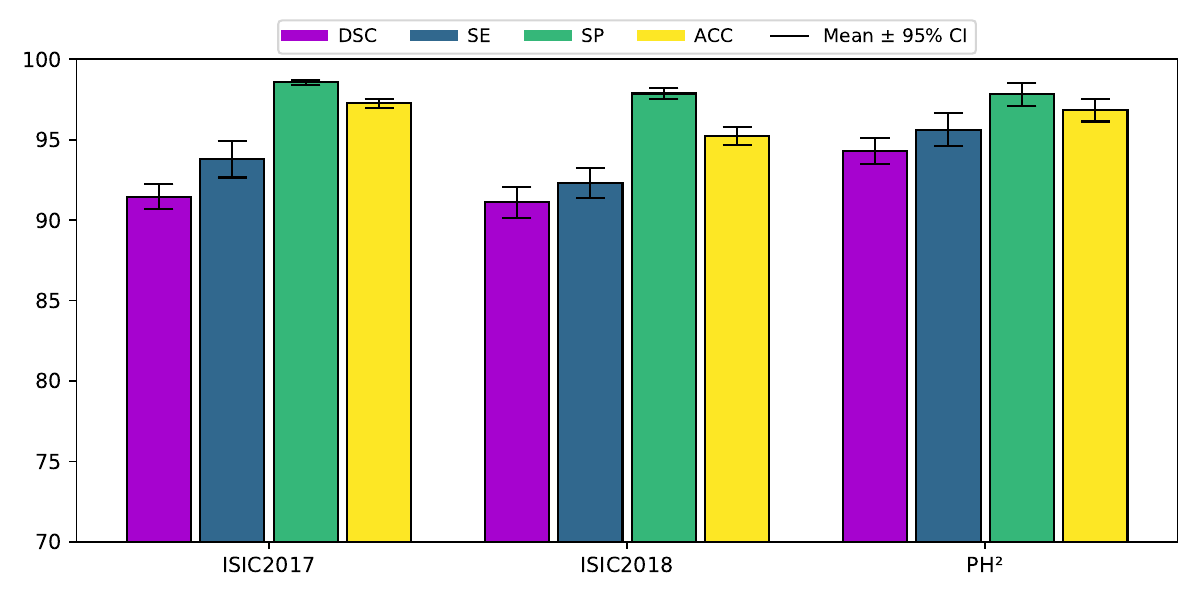}
\end{center}
   \caption{Error bars (95\% confidence interval) of DSC, SE, SP, and ACC on ISIC2017, ISIC2018, and PH$^2$ datasets.}
\label{fig:9}
\end{figure}

\begin{figure}[t!]
\begin{center}
    \includegraphics[width=0.8\linewidth]{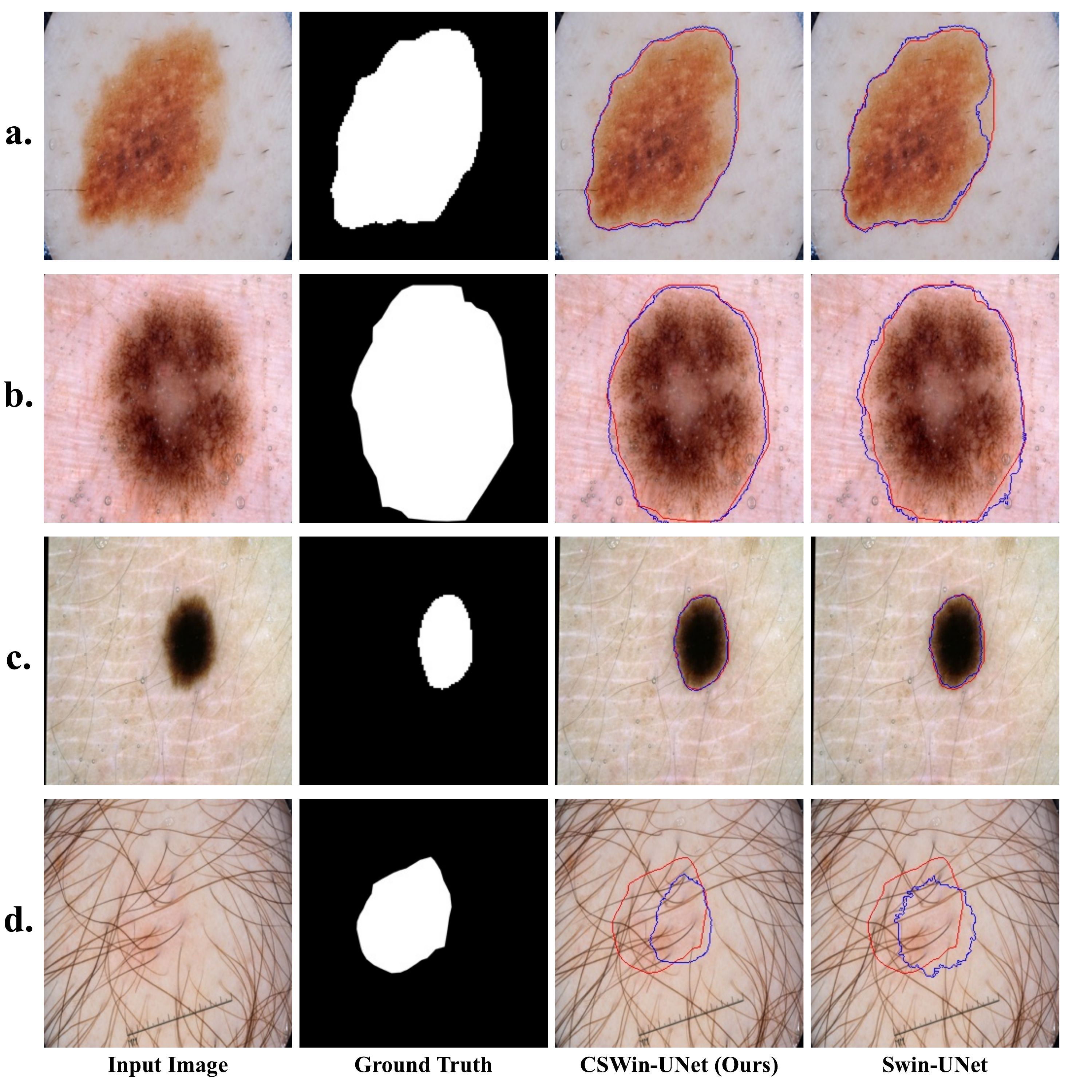}
\end{center}
   \caption{Visual comparison of segmentation results of CSWin-UNet and Swin-UNet on the ISIC2017 dataset. The ground truth and the predicted segmentation boundaries are shown in \textbf{\textcolor[rgb]{1.00,0.00,0.00}{red}} and \textbf{\textcolor[rgb]{0.00,0.00,1.00}{blue}}, respectively.}
\label{fig:8}
\end{figure}

Table \ref{tab:7} shows the experimental results, and Fig.\ref{fig:9} shows the error bars (95\% confidence interval) of DSC, SE, SP, and ACC on three skin lesion segmentation datasets. The experimental results indicate that the proposed CSWin-UNet outperforms other methods in most evaluation metrics. Notably, CSWin-UNet achieves better performance than Swin-UNet\cite{r25} in most metrics, demonstrating satisfactory generalization capability. We also visualized the skin lesion segmentation results in Fig.\ref{fig:8}. Compared to Swin-UNet\cite{r25}, our CSWin-UNet has certain advantages in preserving the edges and detailed features of the segmented objects. However, in cases of low contrast or occlusion, as shown in Fig.\ref{fig:8}(d), the segmentation produces significant errors.

\subsection{Comparison of computational efficiency}

\begin{table}[t!]
\centering
\caption{Computational efficiency of different medical image segmentation methods on the Synapse dataset. The first and second best values are highlighted in \textbf{\textcolor[rgb]{1.00,0.00,0.00}{red}} and \textbf{\textcolor[rgb]{0.00,0.00,1.00}{blue}} fonts, respectively.}
\label{tab:3}
\begin{tabular}{l|c|cc|cc}
\toprule
Model      & Backbone & \#Params (M) & FLOPs (G) & DSC$\uparrow$   & HD$\downarrow$    \\
\midrule
TransUNet\cite{r23} & CNN+Transformer & 96.07        & 88.91     & 77.48 & 31.69 \\
UNetR\cite{r31}      & CNN+Transformer & 86.00        & 54.70     & 79.56 & 22.97 \\
HiFormer-B\cite{r33} & CNN+Transformer & \textbf{\textcolor[rgb]{0.00,0.00,1.00}{25.51}}        & 8.05      & \textbf{\textcolor[rgb]{0.00,0.00,1.00}{80.39}} & \textbf{\textcolor[rgb]{1.00,0.00,0.00}{14.70}} \\
Swin-UNet\cite{r25}  & Transformer & 27.17        & \textbf{\textcolor[rgb]{0.00,0.00,1.00}{6.16}}      & 79.13 & 21.55 \\
\hline
CSWin-UNet & Transformer & \textbf{\textcolor[rgb]{1.00,0.00,0.00}{23.57}}        & \textbf{\textcolor[rgb]{1.00,0.00,0.00}{4.72}}      & \textbf{\textcolor[rgb]{1.00,0.00,0.00}{81.12}} & \textbf{\textcolor[rgb]{0.00,0.00,1.00}{18.86}} \\
\bottomrule
\end{tabular}%
\end{table}

An essential objective for neural network model design is to reduce the parameter count and computational complexity as much as possible while maintaining its performance. This reduction is crucial for enabling more efficient model training and deployment in devices with limited computational resources. Therefore, in evaluating a model, it is necessary to consider not only its accuracy and generalizability but also its parameter count and computational complexity. Here, we use FLOPs and parameters (in millions, M) to measure computational complexity. The performance comparison on the Synapse dataset is shown in Table \ref{tab:3}. The results demonstrate that the proposed CSWin-UNet achieves excellent segmentation performance under conditions of the lowest complexity.

\subsection{Ablation studies}

In this section, we conducted an ablation study on the performance of CSWin-UNet on the Synapse dataset. Specifically, we explored the effects of different upsampling strategies in the decoder, the number of skip connections, different network architecture, and different hyperparameters in the combined loss function on the performance.

\subsubsection{Upsampling strategy}

\begin{table}[t!]
\centering
\caption{Ablation study of different upsampling strategies on the Synapse dataset. The first and second best values are highlighted in \textbf{\textcolor[rgb]{1.00,0.00,0.00}{red}} and \textbf{\textcolor[rgb]{0.00,0.00,1.00}{blue}} fonts, respectively.}
\label{tab:4}
\resizebox{\textwidth}{!}{%
\begin{tabular}{l|c|cccccccc|cc}
\toprule
Strategies &
  DSC $\uparrow$&
  Aorta &
  Gallbladder &
  Kidney (L) &
  Kidney (R) &
  Liver &
  Pancreas &
  Spleen &
  Stomach &
  \#Params(M) &
  FLOPs(G) \\
\midrule
\begin{tabular}[c]{@{}l@{}}Bilinear\\ interpolation\end{tabular} &
  78.82 &
  \textbf{\textcolor[rgb]{0.00,0.00,1.00}{85.32}} &
  57.09 &
  82.68 &
  75.14 &
  94.35 &
  \textbf{\textcolor[rgb]{0.00,0.00,1.00}{65.12}} &
  88.86 &
  \textbf{\textcolor[rgb]{1.00,0.00,0.00}{81.98}} &
  \textbf{\textcolor[rgb]{1.00,0.00,0.00}{23.39}} &
  \textbf{\textcolor[rgb]{1.00,0.00,0.00}{4.38}} \\\hline
\begin{tabular}[c]{@{}l@{}}Transposed\\ convolution\end{tabular} &
  \textbf{\textcolor[rgb]{0.00,0.00,1.00}{80.31}} &
  86.06 &
  65.10 &
  \textbf{\textcolor[rgb]{1.00,0.00,0.00}{85.77}} &
  \textbf{\textcolor[rgb]{1.00,0.00,0.00}{79.79}} &
  \textbf{\textcolor[rgb]{0.00,0.00,1.00}{94.58}} &
  61.73 &
  \textbf{\textcolor[rgb]{1.00,0.00,0.00}{89.68}} &
  79.79 &
  26.03 &
  8.88 \\\hline
CARAFE &
  \textbf{\textcolor[rgb]{1.00,0.00,0.00}{81.12}} &
  \textbf{\textcolor[rgb]{1.00,0.00,0.00}{87.13}} &
  \textbf{\textcolor[rgb]{1.00,0.00,0.00}{67.85}} &
  \textbf{\textcolor[rgb]{0.00,0.00,1.00}{83.51}} &
  \textbf{\textcolor[rgb]{0.00,0.00,1.00}{78.53}} &
  \textbf{\textcolor[rgb]{1.00,0.00,0.00}{95.23}} &
  \textbf{\textcolor[rgb]{1.00,0.00,0.00}{65.94}} &
  \textbf{\textcolor[rgb]{0.00,0.00,1.00}{89.05}} &
  \textbf{\textcolor[rgb]{0.00,0.00,1.00}{81.74}} &
  \textbf{\textcolor[rgb]{0.00,0.00,1.00}{23.57}} &
  \textbf{\textcolor[rgb]{0.00,0.00,1.00}{4.72}} \\
\bottomrule
\end{tabular}%
}
\end{table}

In the encoder, downsampling is performed using a convolutional layer with a stride of 2, and correspondingly, upsampling is needed in the decoder to upsample the feature maps, thereby preserving more information. In this paper, we introduce the CARAFE layer to achieve upsampling and increase feature channels, which uses the content of the input features themselves to guide the upsampling process for more accurate and efficient feature reassembly. To verify the effectiveness of the CAREFE layer, we conducted experiments on the Synapse dataset with bilinear interpolation, transposed convolution, and CARAFE layer in CSWin-UNet, as shown in Table \ref{tab:4}. Upsampling with the CAREFE layer achieved the highest segmentation accuracy. Moreover, compared to transposed convolution, CAREFE introduces very little computational overhead. Experimental results indicate that the CSWin-UNet combined with the CARAFE layer can achieve optimal performance.

\subsubsection{Skip connection}

\begin{table}[t!]
\centering
\caption{Ablation study of different numbers of skip connection on the Synapse dataset. The first and second best values are highlighted in \textbf{\textcolor[rgb]{1.00,0.00,0.00}{red}} and \textbf{\textcolor[rgb]{0.00,0.00,1.00}{blue}} fonts, respectively.}
\label{tab:5}
\resizebox{\textwidth}{!}{%
\begin{tabular}{c|c|cccccccc}
\toprule
\# Skip connection & DSC $\uparrow$ & Aorta & Gallbladder & Kidney (L) & Kidney (R) & Liver & Pancreas & Spleen & Stomach \\
\midrule
0 & 65.41 & 62.22 & 49.25 & 69.89 & 66.60 & 90.21 & 39.10 & 77.57 & 68.42 \\
1 & 76.19 & 84.53 & \textbf{\textcolor[rgb]{0.00,0.00,1.00}{65.51}} & 77.18 & 71.50 & 93.75 & 54.32 & 86.75 & 75.98 \\
2 & \textbf{\textcolor[rgb]{0.00,0.00,1.00}{79.85}} & \textbf{\textcolor[rgb]{0.00,0.00,1.00}{86.21}} & 64.98 & \textbf{\textcolor[rgb]{0.00,0.00,1.00}{81.60}} & \textbf{\textcolor[rgb]{0.00,0.00,1.00}{76.10}} & \textbf{\textcolor[rgb]{0.00,0.00,1.00}{94.27}} & \textbf{\textcolor[rgb]{0.00,0.00,1.00}{62.59}} & \textbf{\textcolor[rgb]{1.00,0.00,0.00}{90.66}} & \textbf{\textcolor[rgb]{1.00,0.00,0.00}{82.44}} \\
3 & \textbf{\textcolor[rgb]{1.00,0.00,0.00}{81.12}} & \textbf{\textcolor[rgb]{1.00,0.00,0.00}{87.13}} & \textbf{\textcolor[rgb]{1.00,0.00,0.00}{67.85}} & \textbf{\textcolor[rgb]{1.00,0.00,0.00}{83.51}} & \textbf{\textcolor[rgb]{1.00,0.00,0.00}{78.53}} & \textbf{\textcolor[rgb]{1.00,0.00,0.00}{95.23}} & \textbf{\textcolor[rgb]{1.00,0.00,0.00}{65.94}} & \textbf{\textcolor[rgb]{0.00,0.00,1.00}{89.05}} & \textbf{\textcolor[rgb]{0.00,0.00,1.00}{81.74}} \\
\bottomrule
\end{tabular}%
}
\end{table}

Similar to the UNet, we also introduced skip connections to enhance finer segmentation details by restoring low-level spatial information. In CSWin-UNet, skip connections are positioned at the resolution scales of 1/4, 1/8, and 1/16. We sequentially reduced the skip connections at the scales of 1/16, 1/8, and 1/4, setting the number of skip connections to 3, 2, 1, and 0 to explore the impact of varying numbers of skip connections on segmentation accuracy. As shown in Table \ref{tab:5}, segmentation accuracy generally improves with an increase in the number of skip connections. Notably, CSWin-UNet achieved more significant improvements in segmentation accuracy on smaller organs (such as the aorta, gallbladder, kidneys, and pancreas) than on larger organs (such as the liver, spleen, and stomach). Therefore, to achieve optimal performance, we set the number of skip connections to 3.

\subsubsection{Network architecture}

\begin{table}[t!]
\centering
\caption{Ablation study of different numbers of CSWin Transformer blocks of each stage on the Synapse dataset. The first and second best values are highlighted in \textbf{\textcolor[rgb]{1.00,0.00,0.00}{red}} and \textbf{\textcolor[rgb]{0.00,0.00,1.00}{blue}} fonts, respectively.}
\label{tab:6}
\begin{tabular}{l|cc|cc}
\toprule
\# Blocks       & \# Params (M) & FLOPs (G) & DSC$\uparrow$   & HD$\downarrow$    \\
\midrule
{[}1, 2, 6, 1{]}  & \textbf{\textcolor[rgb]{1.00,0.00,0.00}{18.82}}        & \textbf{\textcolor[rgb]{1.00,0.00,0.00}{3.79}}      & 79.19 & 24.88 \\
{[}1, 2, 9, 1{]}  & \textbf{\textcolor[rgb]{0.00,0.00,1.00}{23.57}}        & \textbf{\textcolor[rgb]{0.00,0.00,1.00}{4.72}}      & \textbf{\textcolor[rgb]{1.00,0.00,0.00}{81.12}} & \textbf{\textcolor[rgb]{1.00,0.00,0.00}{18.86}} \\
{[}1, 2, 12, 1{]} & 28.32        & 5.65      & \textbf{\textcolor[rgb]{0.00,0.00,1.00}{80.15}} & \textbf{\textcolor[rgb]{0.00,0.00,1.00}{20.43}} \\
\bottomrule
\end{tabular}%
\end{table}

Neural networks with too few layers can result in insufficiently rich and accurate feature representations, making it difficult to understand the image context, thereby leading to poor segmentation performance. Conversely, having too many layers increases the computational burden and makes it difficult for the network to converge. Therefore, when designing the network architecture, a balance was struck between network depth and model performance, enabling the model to achieve high segmentation accuracy with limited computational resources. Additionally, to prevent non-convergence issues due to excessive depth\cite{r35}, the block count in the final stage is set to one. By comparing the number of parameters and computational costs of other Transformer-based medical image segmentation methods, we set the block numbers in the four stages to [1, 2, 6, 1], [1, 2, 9, 1], and [1, 2, 12, 1], with encoder and decoder blocks symmetrically arranged. As shown in Table \ref{tab:6}, the network architecture with block settings of [1, 2, 9, 1] achieved the best performance.

\subsubsection{Combined loss function}\label{sec:4-7-4}

\begin{table}[t!]
\centering
\caption{Ablation study of different hyperparameters of the combined loss function on the Synapse dataset. The first and second best values are highlighted in \textbf{\textcolor[rgb]{1.00,0.00,0.00}{red}} and \textbf{\textcolor[rgb]{0.00,0.00,1.00}{blue}} fonts, respectively.}
\label{tab:8}
\begin{tabular}{cc|cc}
\toprule
$\alpha$ & $\beta$       & DSC$\uparrow$   & HD$\downarrow$    \\
\midrule
1 & 0      & 76.18     & 33.97 \\
0 & 1      & 80.28     & 29.19 \\
0.5 & 0.5  & \textbf{\textcolor[rgb]{0.00,0.00,1.00}{80.79}}     & \textbf{\textcolor[rgb]{0.00,0.00,1.00}{25.65}} \\
0.4 & 0.6  & \textbf{\textcolor[rgb]{1.00,0.00,0.00}{81.12}}     & \textbf{\textcolor[rgb]{1.00,0.00,0.00}{18.86}} \\
0.6 & 0.4  & 80.27     & 27.55 \\
\bottomrule
\end{tabular}%
\end{table}

We explored the impact of different hyperparameters of the combined loss function on segmentation accuracy. Here, we set $\alpha$ and $\beta$ in Eq.\ref{eq:10} to $[1,0]$, $[0,1]$, $[0.5,0.5]$, $[0.4,0.6]$, and $[0.3,0.7]$. We conducted an ablation study on the Synapse dataset, and the experimental results showed that using the combined loss function resulted in higher segmentation accuracy than using either Dice or cross-entropy losses alone, especially in the case of using only Dice loss without cross-entropy loss. Table \ref{tab:8} shows that the segmentation performance was optimal when $\alpha$ and $\beta$ were set to $[0.4,0.6]$.

\subsection{Discussions}

Our comprehensive experimental results on three different types of medical image segmentation datasets, including CT, MRI, and skin lesion images, demonstrate that the proposed CSwin-UNet is more advanced and suitable for medical images of various modalities than other state-of-the-art medical image segmentation methods. However, our method shows some deficiencies in some challenging cases, such as significant differences in segmentation accuracy for different samples of the gallbladder and kidney regions in the Synapse dataset, as shown in Fig.\ref{fig:4}. According to the visualization results in Fig.\ref{fig:8}, there is still much room for improvement in segmentation performance when dealing with low-contrast images in the skin lesion segmentation datasets. Additionally, the pre-training of the model dramatically impacts its performance. In our study, we initialized the encoder and decoder with the weights trained by the CSwin Transformer\cite{r20} on ImageNet\cite{r38}. Therefore, exploring end-to-end medical image segmentation methods is one of the research topics we are striving to pursue in the future.

\section{Conclusion}\label{sec:5}

In this paper, we address the limitations of receptive field interactions in previous Transformer-based medical image segmentation models by introducing an efficient and lightweight method, CSWin-UNet. Utilizing the CSWin self-attention mechanism from the CSWin Transformer, we incorporate this technology into a U-shaped encoder-decoder architecture. This integration not only reduces computational costs but also improves receptive field interactions and segmentation accuracy. In the decoder, the CARAFE layer is employed for upsampling, which helps retain intricate details and enhances the precision of organ edge segmentation. Comprehensive evaluations on three large-scale medical image segmentation datasets illustrate that CSWin-UNet surpasses other state-of-the-art methods in segmentation accuracy. Furthermore, CSWin-UNet is more lightweight regarding model parameters and computational load, suggesting significant potential for further optimizations and enhancements in deep learning applications for complex medical image segmentation tasks.

\bibliographystyle{num}
\bibliography{bibliography}

\end{document}